\documentclass{IEEEtran}
\usepackage{cite}
\usepackage{amsmath,amssymb,amsfonts}
\usepackage{algorithmic}
\usepackage{graphicx}
\usepackage{textcomp}

\usepackage{color}
\usepackage{graphicx}% Include figure files
\usepackage{epstopdf}
\usepackage{dcolumn}% Align table columns on decimal point
\usepackage{bm}% bold math
\usepackage{textcomp}
\usepackage{caption}
\usepackage{subcaption}
\usepackage{array}
\usepackage{tabularx}
\usepackage{multirow}
\usepackage{gensymb}
\usepackage{fancyhdr}
\usepackage{float}
\usepackage{mathtools, cuted} %long euqation break the columns
\usepackage{mathtools} %long euqation break the columns

\def\BibTeX{{\rm B\kern-.05em{\sc i\kern-.025em b}\kern-.08em
    T\kern-.1667em\lower.7ex\hbox{E}\kern-.125emX}}
\begin{document}
\title{Exact solution for surface wave to space wave conversion by periodical impenetrable metasurfaces}
\author{Svetlana~N.~Tcvetkova, \IEEEmembership{Student Member, IEEE}, Stefano Maci \IEEEmembership{Fellow, IEEE}, and Sergei~A.~Tretyakov, \IEEEmembership{Fellow, IEEE}
\thanks{This work was supported in part by the Academy of Finland (project 287894), Nokia Foundation (project 201810155), HPY Research Foundation and European Space Agency. }
\thanks{S.~N.~Tcvetkova and S.~A.~Tretyakov are with the Department of Electronics and Nanoengineering, Aalto University, P.O. 15500, FI-00076 Aalto, Finland  (e-mail: svetlana.tcvetkova@aalto.fi). }
\thanks{S. Maci is with Department of Information Engineering and Mathematics, University of Siena, Via Roma 56, 53100, Siena, Italy.}}

%\author{S.~N.~Tcvetkova}
%\email{svetlana.tcvetkova@aalto.fi}
%\affiliation{Aalto University, Department of Electronics and Nanoengineering, P.O. 15500, FI-00076 Aalto, Finland}
%\author{S. Maci}
%\affiliation{University of Siena, Department of Information Engineering and Mathematics, Via Roma 56, 53100, Siena, Italy}
%\author{S.~A.~Tretyakov}
%\affiliation{Aalto University, Department of Electronics and Nanoengineering, P.O. 15500, FI-00076 Aalto, Finland}

\maketitle

\begin{abstract}
We show that there exists an exact solution for a lossless and reciprocal periodic surface impedance which ensures full conversion of a single-mode surface wave propagating along the impedance boundary to a single plane wave propagating along a desired direction in free space above the boundary. In contrast to known realizations of leaky-wave antennas, the optimal surface reactance modulation which is found here ensures the absence of evanescent higher-order modes of the Floquet wave expansion of fields near the radiating surface. Thus, all the energy carried by the surface wave is used for launching the single inhomogeneous plane wave into space, without accumulation of reactive energy in higher-order modes. 
The results of the study are expected to be useful for creation of leaky-wave antennas with the ultimate efficiency, and, moreover, can potentially lead to novel
applications, from microwaves to nanophotonics.
\end{abstract}


\section{\label{sec:1}Introduction}

Surface-to-propagating wave conversion (or vice versa) is one of the classical problems in antenna theory and techniques, as well as in plasmonics and nanophotonics. 
Conventionally, such transformation is performed by using leaky-wave antennas \cite{oliner2007,monticone_procieee2015}, which can be realized as periodically modulated open or partially open waveguides as well as other structures with periodic perturbations \cite{tierney_ieeeaps2015, minatti_ieeejap2016c, minatti_ieeejap2016b,martinez-ros_ieeejap2012}.
The field structure in the vicinity of open modulated waveguides or reactive surfaces contains higher-order Floquet harmonics, which store reactive fields, limiting the antenna bandwidth.

%Similarly to the conversion between guided and space propagation modes, the problem of anomalous reflection or refraction is followed by fundamental imperfections leading to inevitable power scattering into undesired directions. The study

It is of fundamental importance to find out if an exact solution for the properties of radiating apertures can be found, such that the surface (assuming that it is infinite) would ideally convert a single surface wave into a single propagating plane wave. 
Some attempts to achieve perfect conversion between space and guided modes are known.
A periodic metasurface engineered using the generalized law of reflection~\cite{yu_science2011}, which leads to the requirement of  a linear reflection phase is reported in \cite{sun_natmater2012}. Power conversion efficiency of nearly  100\% was claimed in that paper.
However, the surface wave is not an eigensolution
of the designed metasurface, therefore the structure does not support propagation of the surface wave along the surface. 
As a result, the reported infinitely long gradient converter operates as a nearly perfect absorber at the steady state. 
For periodic finite-length supercell-based gradient converters \cite{qu_epl2013} which suffer from the decoupling effect taking
place at the interface between supercells, the power conversion efficiency of 78\% was predicted.
Further, transformation of a propagating wave into a surface wave using passive and lossy 
periodic structures was discused  in \cite{estakhri_prx2016}, but the reported power
conversion efficiency of such metasurface is quite low ($\approx 7\%$).
In \cite{achouri_metamaterials2016, achouri_arxiv2016}, transformation of surface wave from a beam wave and back into a beam wave using transparent metasurfaces was discussed. However, the propagating beam launched by the structure appears to be
generated by the active constituents of the metasurface instead of being converted from the surface wave propagating along the metasurface.
In \cite{Maci_eff}, the efficiency of a modulated metasurface antenna with non-uniform modulation is found for a finite circular-aperture antenna. The conversion is obtained by changing the local attenuation parameter to synthesize a specific shape of the aperture amplitude distribution. This allows practical aperture efficiency of 80\%, including control of the spill-over effect at the edge.

Recently, new means for the power conversion efficiency enhancement have been proposed for the case of homogeneous plane wave to surface wave conversion using metasurfaces \cite{tcvetkova_prb2018}. The authors assumed a TE-polarized homogeneous normally incident plane wave as an illumination and transformed it into one TM-polarized surface wave with a linear power growth along the surface. Converting the incident propagating wave into an orthogonally polarized surface wave helps to minimize generation of unwanted evanescent Floquet harmonics, but among the found solutions the only solution of the boundary problem which satisfies Maxwell's equations exactly (still imposing essential limitations on the size of the radiating structure) is nonreciprocal and active/lossy. The authors claimed that the most desirable lossless and reciprocal solution can be found approximately, and the power conversion efficiency in this case can reach up to 95\% for a $20\lambda$-long surface.

In this paper, we show that in fact an exact solution exists if the single propagating plane wave in space is an inhomogeneous plane wave. Within this assumption, the required local power balance can be realized exactly in the system of only one surface-bound mode and one propagating plane wave. An infinite-area lossless and reciprocal surface with the properties found here operates as an ideal leaky-wave antenna, launching only one plane wave in a given direction, not storing any reactive energy in higher-order Floquet modes. 
 The conversion efficiency is predicted to be theoretically ideal and numerically tested to be above 99\%. The found solution can be considered as a reference solution for modulated impedance leaky-wave antennas. 
Considering radiation from a partially transparent wall of a waveguide, a solution with only one propagating  plane wave in space was recently found in \cite{Ariel}. In this paper, we 
study radiation from a modulated reactive boundary, and our exact solution contains only two plane waves that exchange energy with each other.

\section{\label{sec:2}Problem statement}

Let us assume that an infinite lossless boundary (defined by its surface impedance) is illuminated by a plane wave, for example, at normal incidence. Our goal is engineering the surface impedance of the boundary so that all the incident power carried by the plane wave is converted into a single surface wave maintained by the impedance boundary and propagating into a desired direction. While we study  the limiting case of an infinite surface, neglecting scattering by the edges, in practice, the size of the receiving surface is finite and the received energy is collected by a matched load at the edge of the receiving surface. 

%leaky-wave propagating in space.

There are two main difficulties in realizing ideal converters.
Similarly to the anomalous reflectors, one of the problems is to avoid active regions on impenetrable surfaces \cite{asadchy_prb2016}. Because the propagating  plane wave and the surface wave in general interfere,  the required scalar surface impedance has both real and imaginary parts, which corresponds to a  lossless in the average, but locally active and lossy converting surface. It is known that for anomalous reflectors this problem can be overcome if the incident and anomalously reflected waves are orthogonally polarized \cite{asadchy_prb2016}, and we adopt the same approach for the thought ideal converter between propagating and surface waves. 

Another problem is to ensure power balance between the power carried by the incident propagating plane wave and the power accepted by the surface mode at every point of the boundary. 
Since we want to realize wave conversion using a locally responding and lossless impedance surface, the normal component of the real part of the total Poynting vector must be equal to zero at every point of the surface.
However, the Poynting vector of the incident plane wave is uniform over the surface, but all possible surface modes supported by periodic impedance boundaries have exponential field attenuation (or growth), which does not match the distribution of the incident power over the receiving surface.  
Therefore, the exact power balance cannot be satisfied within this scenario (see detailed discussion in \cite{tcvetkova_prb2018}). 
We expect that exact conversion between a single surface mode and a single {\it inhomogeneous} propagating wave may be possible, because in this case the fields of {\it both} surface and space waves have exponential coordinate dependence. Here we explore this possibility and show that an exact solution indeed exists. {\textcolor{black}{We assume that the surface is reciprocal, thus, it can be also seen as converting a surface wave into an inhomogeneous propagating plane wave.}

The ideal conversion scenario (no dissipation or scattering losses) of {\it one} TM-polarized surface wave into a single TE-polarized inhomogeneous plane wave is illustrated in Fig.~\ref{fig:1}. The surface wave is launched along an impenetrable reactive  metasurface characterized by a periodic tensor surface impedance $\overline{\overline {{Z}}} (x)=j\overline{\overline {{X}}} (x)$ with period $d$. 
For simplicity, we assume  that both fields are invariant with respect to the $z$-axis.
The design goal is to find the periodic distribution of $\overline{\overline {{Z}}} (x)$ that allows to perform perfect conversion, while the fields of both waves satisfy Maxwell's equation in the upper half-space (which is considered to be free space).
\begin{figure}[htp]
	\centering
		\includegraphics[width=0.45\textwidth]{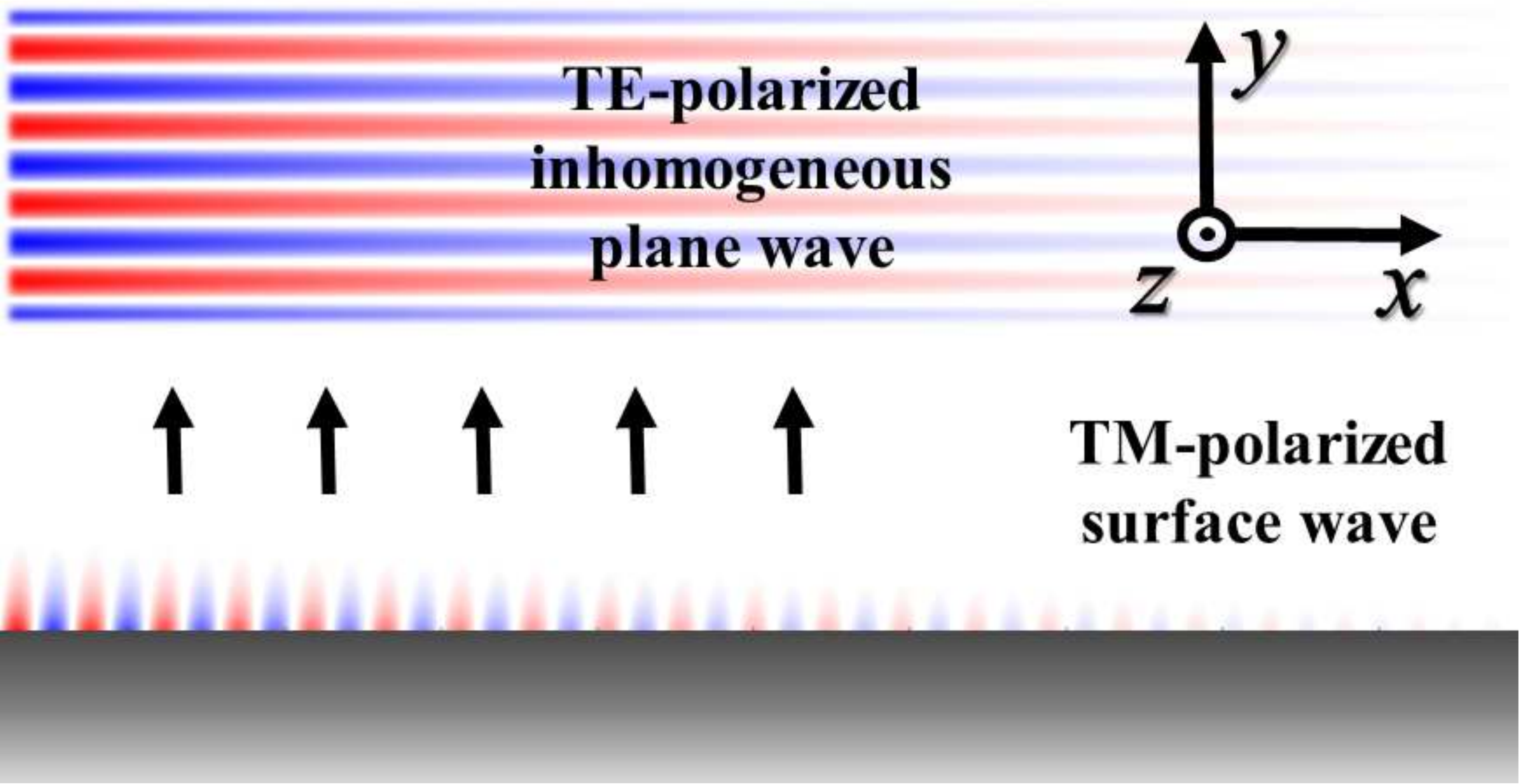}
	\caption{Boundary problem of surface wave to inhomogeneous plane wave conversion.}
	\label{fig:1}
\end{figure}

A complete eigen-mode basis set for the 2D problem in the absence of sources is given by  Floquet-wave modes. The modes are inhomogeneous plane waves with  $x$-components of the complex wavenumbers given by
\begin{equation}
{k_{xn}} = \beta _x - j{\alpha _x} + {\displaystyle{{2\pi n} \over d}},
\end{equation}   	     
where $n$ is the Floquet mode index, $\beta _x$ is the propagating constant, and $\alpha _x$ is the attenuation constant. 
Let us set the surface wave as the $0$-indexed mode of the Floquet expansion. 
The TM-polarized surface wave field components are given by 
\begin{gather}
H_z^{{\rm{sw}}} = H_0^{{\rm{sw}}}{e^{ - \left( {{\alpha _x} + j\beta _x^{{\rm{sw}}}} \right)x}}{e^{ - \left( {\alpha _y^{{\rm{sw}}} + j\beta _y^{{\rm{sw}}}} \right)y}}, \label{eq:sw1}\\
E_x^{{\rm{sw}}} =  - H_0^{{\rm{sw}}}\frac{{\left( {\beta _y^{{\rm{sw}}} - j\alpha _y^{{\rm{sw}}}} \right){\eta _0}}}{{{k_0}}}{e^{ - \left( {{\alpha _x} + j\beta _x^{{\rm{sw}}}} \right)x}}{e^{ - \left( {\alpha _y^{{\rm{sw}}} + j\beta _y^{{\rm{sw}}}} \right)y}}, \label{eq:sw2} \\
E_y^{{\rm{sw}}} = H_0^{{\rm{sw}}}\frac{{\left( {\beta _x^{{\rm{sw}}} - j{\alpha _x}} \right){\eta _0}}}{{{k_0}}}{e^{ - \left( {{\alpha _x} + j\beta _x^{{\rm{sw}}}} \right)x}}{e^{ - \left( {\alpha _y^{{\rm{sw}}} + j\beta _y^{{\rm{sw}}}} \right)y}}.
\label{eq:sw3}
\end{gather}  
The components of the complex wavenumber are 
\begin{gather}
{k_{x}^{{\rm{sw}}}} = \beta _x^{{\rm{sw}}} - j{\alpha _x} ,\\
{k_{y}^{{\rm{sw}}}} = \beta _y^{{\rm{sw}}} - j{\alpha _y^{{\rm{sw}}} } ,
\end{gather}
where $\beta _x^{{\rm{sw}}}$ and $\beta _y^{{\rm{sw}}}$ are the  propagation constants of the surface wave, $\alpha _x$ and $\alpha _y^{\rm sw}$ are the attenuation constants of the surface wave ($x$- and $y$- components, respectively). 
Complex wavenumbers satisfy the free-space dispersion relation
\begin{equation}
{\left( {\beta _x^{\rm{sw}}} - j{\alpha _x} \right)^2} + {\left( {\beta _y^{\rm{sw}}} - j\alpha _y^{\rm{sw}} \right)^2} = {k_0}^2,
\end{equation}
where $k_0$ is the free-space wavenumber.
Separation of the real and imaginary parts leads to
\begin{gather}
\beta _x^{{\rm{sw}}}{\alpha _x} =  - \beta _y^{{\rm{sw}}}\alpha _y^{{\rm{sw}}},\label{eq:8}\\ 
{\left( {\beta _x^{{\rm{sw}}}} \right)^2} - {\alpha _x}^2 + {\left( {\beta _y^{{\rm{sw}}}} \right)^2} - {\left( {\alpha _y^{{\rm{sw}}}} \right)^2} = {k_0}^2.\label{eq:9}
\end{gather}
Any combination of the values of the four parameters that satisfy (\ref{eq:8}) and (\ref{eq:9}) corresponds to fields which satisfy Maxwell's equations above the boundary, at $y>0$. An evanescent in the $y$-direction and decaying along the $x$-direction wave is associated with the conditions $\beta _x^{{\rm{sw}}}>k_0$ and $\alpha _x>0$. Thus,  $\beta _y^{{\rm{sw}}}<0$ and $\alpha _y^{\rm sw}>0$.

If the impedance modulation period $d$ is smaller than the free-space wavelength at the operation frequency, the propagation constant of the $-1$-indexed mode is smaller than $k_0$, thus the wave becomes a propagating one. In other words, the energy is leaking from the surface. The complex wavenumber of the $-1$-indexed mode is corresponding to a leaky-wave mode, which exponentially decays along the surface:
\begin{equation}
{k_{x}^{{\rm{lw}}}} = \beta _x^{{\rm{sw}}} - j{\alpha _x}- {2\pi \over d}=\beta _x^{{\rm{lw}}} - j{\alpha _x} .
\label{eq:1}
\end{equation}
Here, ${\alpha _x}$ is identical for both the surface and the leaky waves.
The propagation constant of the plane wave depends on the direction of propagation, $\beta_x^{{\rm{lw}}}=k_0 \, {\rm sin} \theta $, where $\theta$ is the angle between the wave propagation direction and the normal to the surface.  \textcolor{black}{ The solution can be found for any incidence angle, however, for simplicity,}
let us consider the case when the phase constant of the leaky wave has negligibly small tangential component: $\beta _x^{{\rm{lw}}} \approx 0$. Therefore, from (\ref{eq:1}), $\beta _x^{{\rm{sw}}} \approx {\displaystyle{{2\pi } / d}}$. 
%the leaky wave can propagate rigorously in the vertical direction only if $\beta _x^{{\rm{lw}}}=0$, $\alpha _y^{{\rm{lw}}} = 0$ and, therefore,
Thus, the TE-polarized leaky wave \textcolor{black}{propagates along the normal} and its field components are expressed as
\begin{gather}
E_z^{{\rm{lw}}} = E_0^{{\rm{lw}}}{e^{ - {\alpha _x}x}}{e^{ - j\beta _y^{{\rm{lw}}}y}},\\
H_x^{{\rm{lw}}} =  E_0^{{\rm{lw}}}\frac{{\beta _y^{{\rm{lw}}}}}{{{\eta _0}{k_0}}}{e^{ - {\alpha _x}x}}{e^{ - j\beta _y^{{\rm{lw}}}y}},\\
H_y^{{\rm{lw}}} = E_0^{{\rm{lw}}}\frac{{j{\alpha _x}}}{{{\eta _0}{k_0}}}{e^{ - {\alpha _x}x}}{e^{ - j\beta _y^{{\rm{lw}}}y}}.
\end{gather}  
with the free-space dispersion relation given by               
\begin{equation}
{\left( {\beta _x^{{\rm{lw}}} - j{\alpha _x}} \right)^2} + {\left( {\beta _y^{{\rm{lw}}} - j\alpha _y^{\rm{lw}}} \right)^2} = {k_0}^2 \end{equation}
or, in case of leakage along the normal direction,
\begin{gather}
{\left( {\beta _y^{{\rm{lw}}}} \right)^2} - {\alpha _x}^2 = {k_0}^2. \label{eq:15}
\end{gather}

The wave coupler as a point-wise lossless impenetrable metasurface can be realized when the superposition of the decaying surface wave and the leaky wave propagating away from the surface satisfy zero net power density penetration in the $xz$-plane everywhere \cite{dhkwon_net, tcvetkova_prb2018}.
Therefore, the {\textcolor{black}{real part of the}} normal component of the total power density is required to be equal zero on the boundary ($y$=0): 
\begin{equation}
\begin{split}
S_y (x,y=0)=S_y^{{\rm{sw}}} + S_y^{{\rm{lw}}}& = {1\over 2}{\beta_y^{\rm sw} \eta_0\over k_0} |H_0^{\rm sw}|^2 e^{-2\alpha_x x} \\& +{1\over 2}{\beta_y^{\rm lw}\over k_0 \eta_0} |E_0^{\rm lw}|^2 e^{-2\alpha_x x}=0,
\end{split}
\end{equation}
where $S_y^{{\rm{sw}}}$ and $S_y^{{\rm{lw}}}$ are the normal components of the Poynting vector associated with the surface wave and leaky wave, respectively.
This condition defines the following magnetic field magnitude of the surface wave:
\begin{equation} 
|H_0^{{\rm{sw}}}| = \frac{|{E_0^{{\rm{lw}}}}|}{{{\eta _0}}}\sqrt {-\frac{{ {\beta _y^{{\rm{lw}}}} }}{{\beta _y^{{\rm{sw}}}}}} ,  
\label{eq:H0mag}\end{equation}
where $\beta _y^{{\rm{sw}}}$ is a negative value, as mentioned above.
Fulfilling this condition ensures that all the power from the surface wave is entirely converted in the single radiated leaky wave. Reciprocally, the power density transported by the incident propagating wave with the same parameters is entirely received by the surface wave.

\section{\label{sec:4}Solution and results}

To find the type of the surface needed for ideal conversion, the impedance boundary condition at $y$=0 is considered.
The total tangential fields on the surface are given by 
\begin{equation}
\begin{split}
{{\bf{H}}_{\rm{t}}} = {\bf{z}}H_{{\rm t}z}+{\bf{x}}H_{{\rm t}x}={\bf{z}}H_0^{{\rm{sw}}}&{e^{ - \left( {{\alpha _x} + j\beta _x^{{\rm{sw}}}} \right)x}}\\& +{\bf{x}}E_0^{{\rm{lw}}}\frac{{\beta _y^{{\rm{lw}}}}}{{{\eta _0}{k_0}}}{e^{ - {\alpha _x}x}},
\end{split}
\label{eq:Ht}
\end{equation}
\begin{equation}
\begin{split}
{{\bf{E}}_{\rm{t}}} = {\bf{z}}E_{{\rm t}z}&+{\bf{x}}E_{{\rm t}x}={\bf{z}}E_0^{{\rm{lw}}}{e^{ - {\alpha _x}x}} \\&- {\bf{x}}H_0^{{\rm{sw}}}\frac{{\left( {\beta _y^{{\rm{sw}}} - j\alpha _y^{{\rm{sw}}}} \right){\eta _0}}}{{{k_0}}}{e^{ - \left( {{\alpha _x} + j\beta _x^{{\rm{sw}}}} \right)x}},
\end{split}
\label{eq:Et}
\end{equation}
where $E_{{\rm t}z}$, $E_{{\rm t}x}$ and $H_{{\rm t}z}$, $H_{{\rm t}x}$ are the tangential components of the electric and magnetic fields, respectively.

The surface located at $y=0$ can be characterized by a surface impedance tensor $\overline{\overline {Z}}$, which relates the total tangential electric and magnetic fields as 
\begin{equation}
{\bf {E}}_{\rm{t}}^{} = \overline{\overline {{Z}}}  \cdot \left( {{\bf{y}} \times {{\bf{H}}_{\rm{t}}}} \right) = \left[ {\begin{array}{*{20}{c}}
{{Z_{xx}}}&{{Z_{xz}}}\\
{{Z_{zx}}}&{{Z_{zz}}}
\end{array}} \right] \cdot \left( {{\bf{y}} \times {{\bf{H}}_{\rm{t}}}} \right),
\label{eq:tenzor_rel} \end{equation}                   
where ${Z_{xx}}$, ${Z_{xz}}$, ${Z_{zx}}$, and ${Z_{zz}}$ are components of the anisotropic impedance tensor, which generally can be complex numbers. 
Since there are two complex-valued equations in (\ref{eq:tenzor_rel}) for four complex-valued impedance elements, the solution is not unique. 
Thus, it gives some freedom to choose the values of the impedance elements. 
The most desirable and practically implementable case of lossless and reciprocal system is considered further. 
All the elements of the impedance matrix for such boundaries are purely imaginary and the matrix is skew-Hermitian: ${Z_{xx}} = j{X_{xx}}$, ${Z_{xz}} = {Z_{zx}} = j{X_{xz}}$, and ${Z_{zz}} = j{X_{zz}}$. 
 \begin{figure}[htp]
	\centering
	 \begin{subfigure}[h]{0.47\textwidth}
        \includegraphics[width=\textwidth]{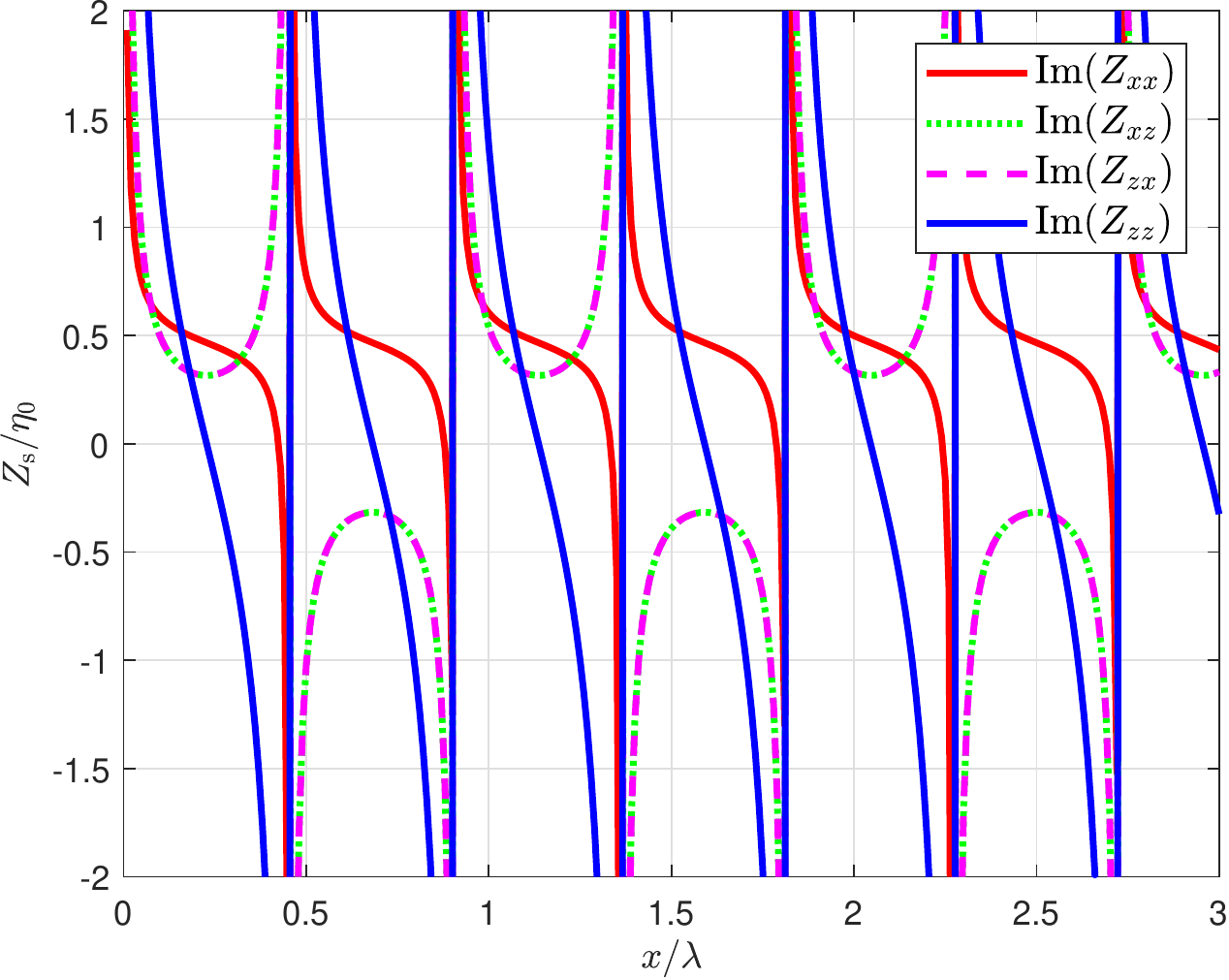}
    \end{subfigure}\\  
	\caption{(a) Surface impedance normalized to the free space for the reciprocal and lossless boundary for the wave solution with ${\alpha _x} =  0.0425{k_0}$, $\alpha _y^{{\rm{sw}}} = 0.4671{k_0}$, $\beta _x^{{\rm{sw}}} = 1.1{k_0}$,$\beta _y^{{\rm{sw}}} = -0.1{k_0}$, $\beta _x^{{\rm{lw}}}=0$, $\beta _y^{{\rm{lw}}} =1.0009 {k_0}$, and $\alpha _y^{\rm lw}=0$. }
	\label{fig:Zs}
\end{figure}
Inspection of (\ref{eq:Ht}), (\ref{eq:Et}) along with the use of (\ref{eq:H0mag}) allow to recognize the four components of the impedances as
\begin{equation}
\overline{\overline {{Z}}}  = j\left[ {\begin{array}{*{20}{c}}
{ \displaystyle\frac{{{\eta _0}}}{{{k_0}}}\left( {\alpha _y^{{\rm{sw}}} - \beta _y^{{\rm{sw}}}\cot \beta _x^{{\rm{sw}}}x} \right)}&{\displaystyle\frac{{{\eta _0}}}{{\sin \beta _x^{{\rm{sw}}}x }}\sqrt {-\frac{{\beta _y^{{\rm{sw}}}}}{{ {\beta _y^{{\rm{lw}}}}}}}}\\
{\displaystyle\frac{{{\eta _0}}}{{\sin \beta _x^{{\rm{sw}}}x}}\sqrt {-\frac{{\beta _y^{\rm{sw}}}}{ {\beta _y^{\rm{lw}}} }}}&{\displaystyle \frac{{{k_0}{\eta _0}}}{ {\beta _y^{{\rm{lw}}}} }\cot \beta _x^{{\rm{sw}}}x{\rm{ }}}
\end{array}} \right].
\label{eq:Zs}
\end{equation}
Figure~\ref{fig:Zs} shows the surface reactance values  $\overline{\overline {{X}}} $ for some chosen values of the wave parameters. \textcolor{black}{The surface impedance in case of oblique incidence is presented in Appendix \ref{ap:1}.}

We note that if the impedance boundary conditions for the reciprocal and lossless system are satisfied by one surface wave and a single leaky wave, the exact solution of the boundary value problem for an impedance surface $j\overline{\overline {{X}}}$ does not include any higher-order Floquet modes of the field expansion. 
In the following, a numerical verification is presented.

\subsection{\label{sec:4.1}	Visualization of the exact solution }
\begin{figure}[H]
	\centering
	\vspace{-0.5cm}		\begin{gather*}{
         \includegraphics[width=0.35\textwidth]{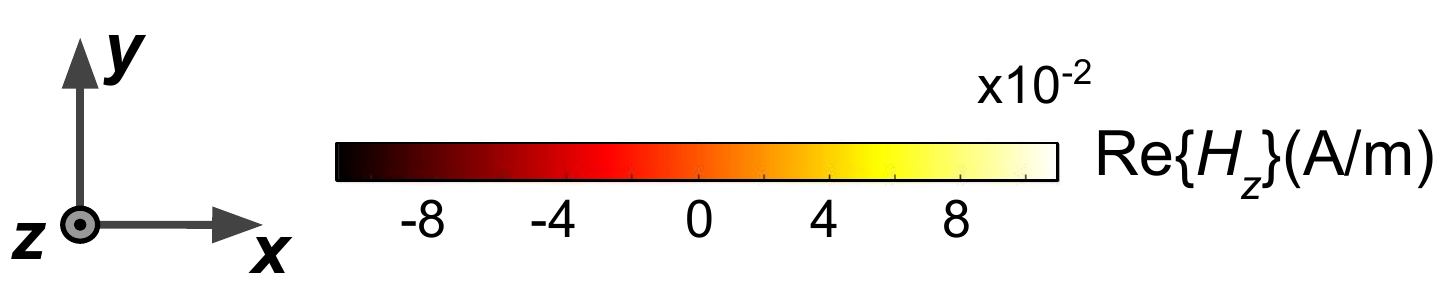}}               
              \end{gather*}  
\begin{subfigure}[h]{0.47\textwidth}
     	\vspace{-0.3cm}    \includegraphics[width=\textwidth]{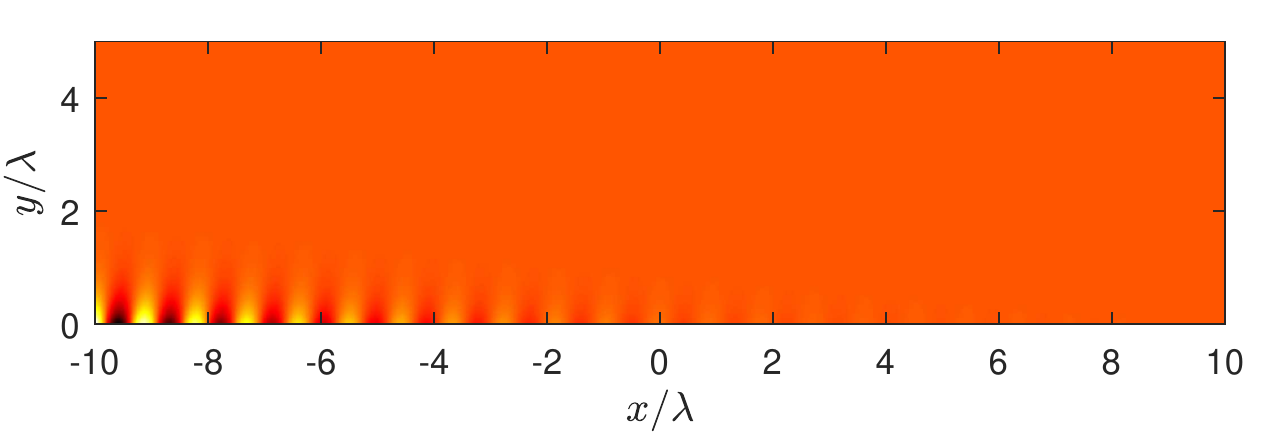}
        \caption{}
    \end{subfigure}\\  
    		\begin{gather*}{
        \includegraphics[width=0.35\textwidth]{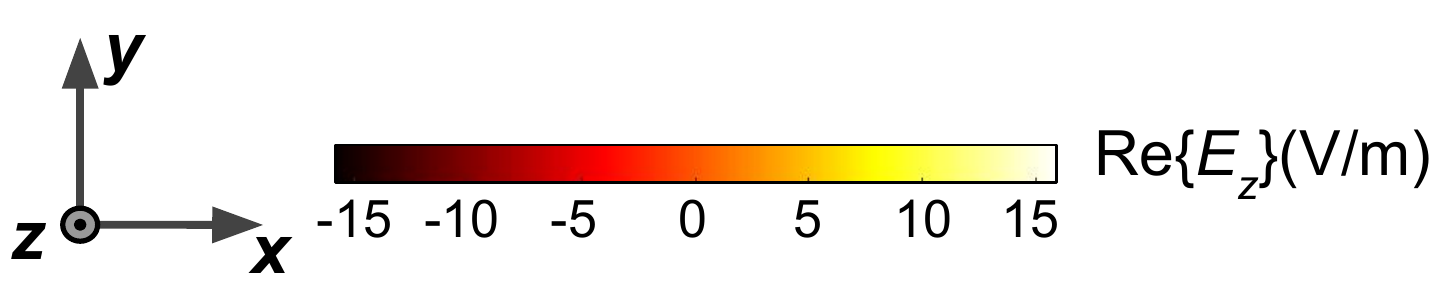}}               
              \end{gather*}  
     \begin{subfigure}[h]{0.47\textwidth}
      	\vspace{-0.3cm}   \includegraphics[width=\textwidth]{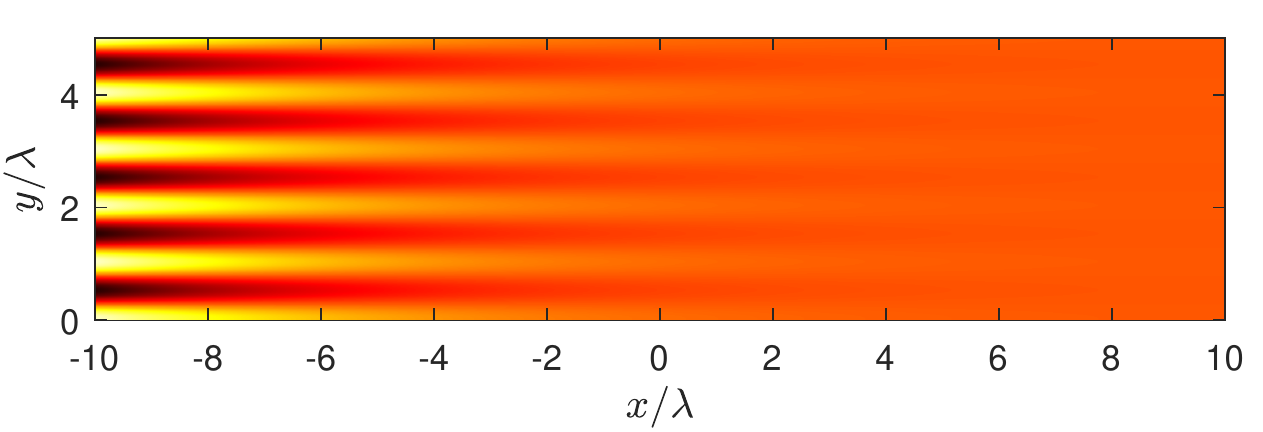}
        \caption{}
    \end{subfigure}
	\caption{Ideal distributions of (a) magnetic field ${\rm{Re}}(H_z) $ of the TM-polarized single surface wave and (b) electric field ${\rm Re}(E_z)$ of the TE-polarized inhomogeneous plane wave with the following wave parameters: ${\alpha _x} =  0.0425{k_0}$, $\alpha _y^{{\rm{sw}}} = 0.4671{k_0}$, $\beta _x^{{\rm{sw}}} = 1.1{k_0}$,$\beta _y^{{\rm{sw}}} = -0.1{k_0}$, $\beta _x^{{\rm{lw}}}=0$, $\beta _y^{{\rm{lw}}} =1.0009 {k_0}$, and $\alpha _y^{\rm lw}=0$.}
	\label{fig:4.1}
\end{figure}
We consider a rectangular cross-section area in the $xy$-plane with the length $L = 20\lambda$ and the height $H = 5\lambda$, where $\lambda$ is the free-space wavelength. Impenetrable impedance boundary condition is used to model the surface by impressing
%\noindent 
electric currents specified in terms of electric fields and a surface admittance matrix. The propagating constants of the surface wave for the case of $20\lambda$-long surface are chosen to be $\beta _x^{{\rm{sw}}} = 1.1{k_0}$, $\beta _y^{{\rm{sw}}} = -0.1{k_0}$, therefore, from (\ref{eq:8}), (\ref{eq:9}) and (\ref{eq:15}) all other  parameters of the waves can be found, and they read ${\alpha _x} =  0.0425{k_0}$, $\alpha _y^{{\rm{sw}}} = 0.4671{k_0}$, $\beta _x^{{\rm{lw}}}=0$, $\beta _y^{{\rm{lw}}} =1.0009 {k_0}$, and $\alpha _y^{\rm lw}=0$. The operating frequency is $f = 10$~GHz, but the results can be scaled to any frequency.
Figures~\ref{fig:4.1}(a) and (b) visualize the exact field solutions for the $z$-components of the magnetic field of the TM-polarized surface wave and of the electric field of the TE-polarized inhomogeneous plane wave, respectively. The results are obtained using MATLAB \cite{matlab}.

The next step is to verify how close the conversion of the surface wave into a leaky wave using the surface impedance to the ideal picture.

\subsection{\label{sec:4.2}Numerical full-wave analysis }

By using full-wave simulations with COMSOL Multiphysics \cite{comsol}, the conversion efficiency \cite{Maci_eff} of the metasurface can be found. The parameters are identical to the MATLAB simulations presented above. On the top and on the right 
\begin{figure}[H]
	\centering
	\vspace{-0.5cm}		\begin{gather*}{
        \includegraphics[width=0.35\textwidth]{ReHz.pdf}}               
              \end{gather*}  
     \begin{subfigure}[h]{0.47\textwidth}
      	\vspace{-0.2cm}   \includegraphics[width=\textwidth]{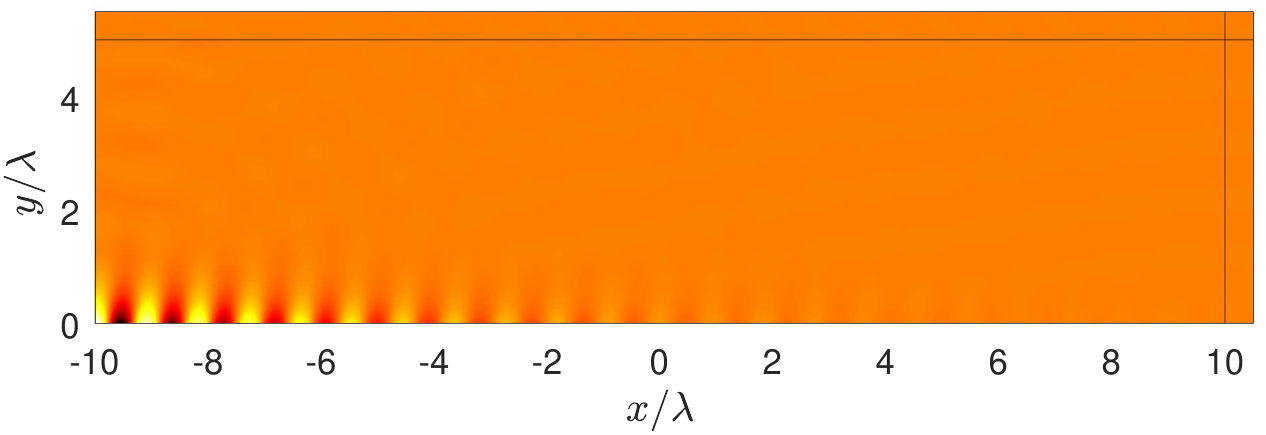}
        \caption{}
    \end{subfigure}	
			\begin{gather*}{
        \includegraphics[width=0.35\textwidth]{ReEz.pdf}}               
              \end{gather*}  
	 \begin{subfigure}[h]{0.47\textwidth}
      	\vspace{-0.2cm}   \includegraphics[width=\textwidth]{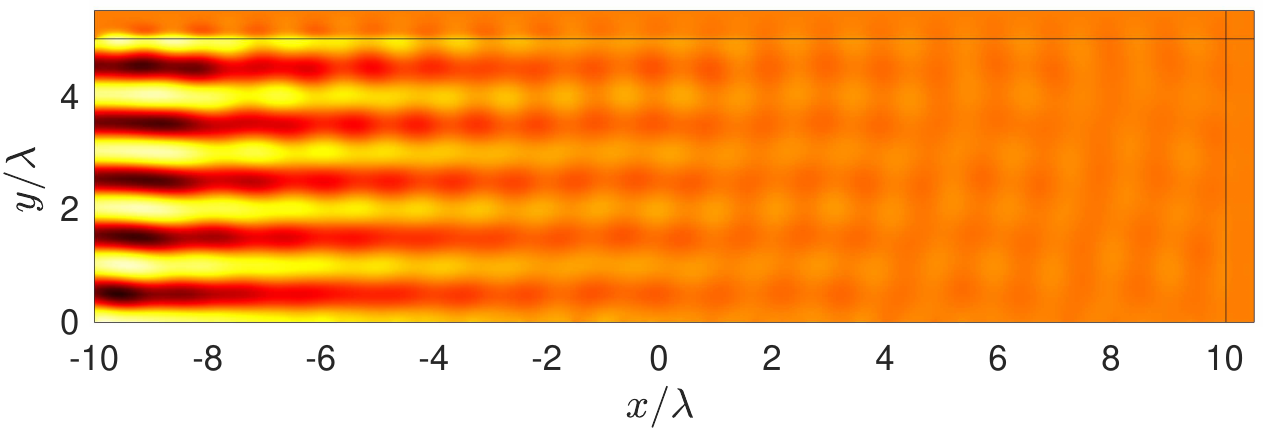}
        \caption{}
    \end{subfigure}
	\caption{\textcolor{black}{Numerically computed distributions of the tangential components of the total (a) magnetic field ${\rm{Re}}(H_z) $ and (b) electric field ${\rm Re}(E_z)$ }with the following wave parameters: ${\alpha _x} =  0.0425{k_0}$, $\alpha _y^{{\rm{sw}}} = 0.4671{k_0}$, $\beta _x^{{\rm{sw}}} = 1.1{k_0}$,$\beta _y^{{\rm{sw}}} = -0.1{k_0}$, $\beta _x^{{\rm{lw}}}=0$, $\beta _y^{{\rm{lw}}} =1.0009 {k_0}$, and $\alpha _y^{\rm lw}=0$.}
	\label{fig:4.2a}
\end{figure} 
\noindent boundaries of the computation domain the perfectly matched layer condition is specified by the box with the  thickness of ${\lambda}/2$. 
Double-port condition is set on the left side of the box to allow the surface wave excitation with the specified settings, while not perturbing the converted inhomogeneous plane wave (\textcolor{black}{the double-port condition is} set in the software manually).
The general mesh of the model has the maximum size of the element equal to ${\lambda}/20$, while finer mesh elements required along the impedance boundary are ${\lambda}/300$. 
Figures~\ref{fig:4.2a}(a) and (b) show the $z$-component of the total electric and magnetic fields, respectively, for the lossless and reciprocal system. 
The conversion efficiency is determined as the ratio of the power carried by the TE-polarized  inhomogeneous plane wave, which is detected at the distance of $5\lambda$ away from the surface to the power launched by the port and carried by TM-polarized surface wave. For the considered case the conversion efficiency of the surface with the length of $20\lambda$ is around $99.8\%$. 

\begin{figure}[H]
	\centering
	\vspace{-0.5cm}		\begin{gather*}{
        \includegraphics[width=0.35\textwidth]{ReHz.pdf}}               
              \end{gather*}  
     \begin{subfigure}[h]{0.47\textwidth}
      	\vspace{-0.2cm}   \includegraphics[width=\textwidth]{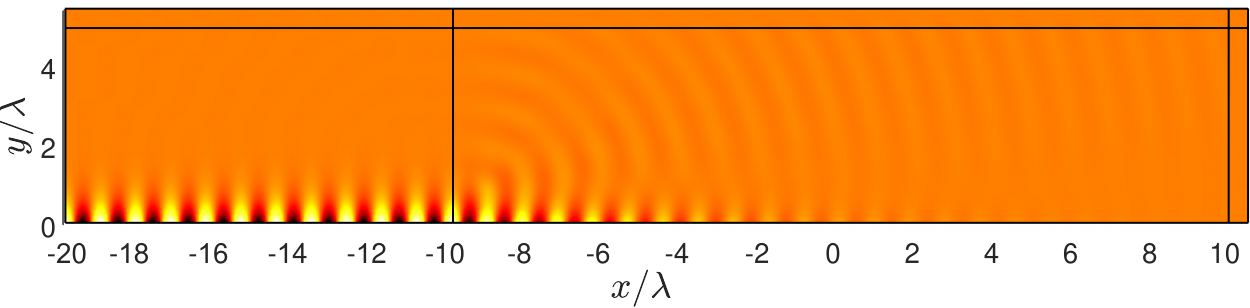}
        \caption{}
    \end{subfigure}	
			\begin{gather*}{
        \includegraphics[width=0.35\textwidth]{ReEz.pdf}}               
              \end{gather*}  
	 \begin{subfigure}[h]{0.47\textwidth}
      	\vspace{-0.2cm}   \includegraphics[width=\textwidth]{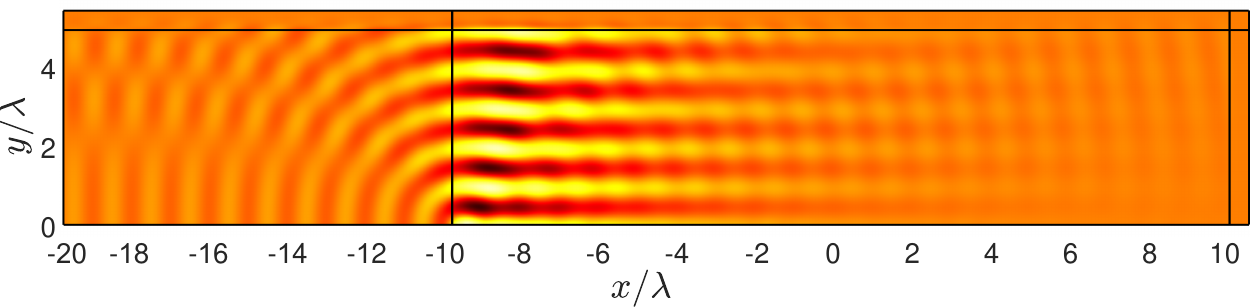}
        \caption{}
    \end{subfigure}
	\caption{\textcolor{black}{Numerically computed distributions of the tangential components of the total (a) magnetic field ${\rm{Re}}(H_z) $ and (b) electric field ${\rm Re}(E_z)$ for the metasurface with augmented waveguide described by (\ref{eq:Xwg}). The wave parameters are: ${\alpha _x} =  0.0425{k_0}$, $\alpha _y^{{\rm{sw}}} = 0.4671{k_0}$, $\beta _x^{{\rm{sw}}} = 1.1{k_0}$,$\beta _y^{{\rm{sw}}} = -0.1{k_0}$, $\beta _x^{{\rm{lw}}}=0$, $\beta _y^{{\rm{lw}}} =1.0009 {k_0}$, and $\alpha _y^{\rm lw}=0$.}}
	\label{fig:wg}
\end{figure}

\textcolor{black}{Many cases of conversion considered in the literature use the surface waves which are not eigensolutions of the surface, but so-called ``driven modes", which cannot propagate along the structure. Therefore, such structures work as nearly perfect absorbers at the steady state. However, in the present work the surface wave can be guided in or out of the system, since it is an eigensolution. In order to confirm this conclusion,  we numerically tested the described converter of a  finite length which is augmented by a homogeneous reactive surface, acting as a feeder surface-wave waveguide. A snapshot of the total magnetic, $H_z$, and electric, $E_z$, fields are represented in Fig.~\ref{fig:4.2a}(a) and (b), respectively. At the location of the original port launching the surface wave at 
$x=-10\lambda$, a reactive surface of length $10\lambda$
is connected. Its surface reactance value, $X_{s,\text{guide}}$, is set to 
\begin{equation}
X_{s,\text{guide}}={\eta_0 \over k_0}\alpha_y^{\rm sw}
\label{eq:Xwg} \end{equation}
to support a TM-polarized surface wave as an eigenwave. A port for launching the surface wave is specified at the beginning of the simulation domain at $x=-20\lambda_0$ and set manually.
Fig.~\ref{fig:wg}(a) shows a snapshot of the tangential component of the total magnetic field, $H_z$, of the metasurface with the augmented structure. It is clear that a surface wave propagates in the waveguide unattenuated as a proper eigenwave. The junction of two different surface impedances causes some  scattering of power ($\approx10\%$) due to not ideal mode and impedance matching of the two surfaces  (the metasurface is designed to support a decaying surface wave (\ref{eq:sw1}-\ref{eq:sw3}) while the waveguide supports a surface wave with a constant amplitude). Fig.~\ref{fig:wg}(b) shows a snapshot of the tangential component of the total electric field, $E_z$, of the metasurface with augmented structure. As  expected, the inhomogeneous plane wave is leaking to the waveguide structure creating ripples. However, power conversion is not compromised: All the energy carried by the surface wave after the junction is received by the inhomogeneous plane wave. } Next, a possibility of shortening the converting surface length is considered.

\subsection{\label{sec:4.3}{Surface length reduction}}
Figure~\ref{fig:4.2b} shows the model with a rectangular cross-section box in the $xy$-plane with the length $L = 6\lambda$ and the height $H = 5\lambda$. 
\begin{figure}[htp]
	\centering
\begin{gather*}{
        \includegraphics[width=0.35\textwidth]{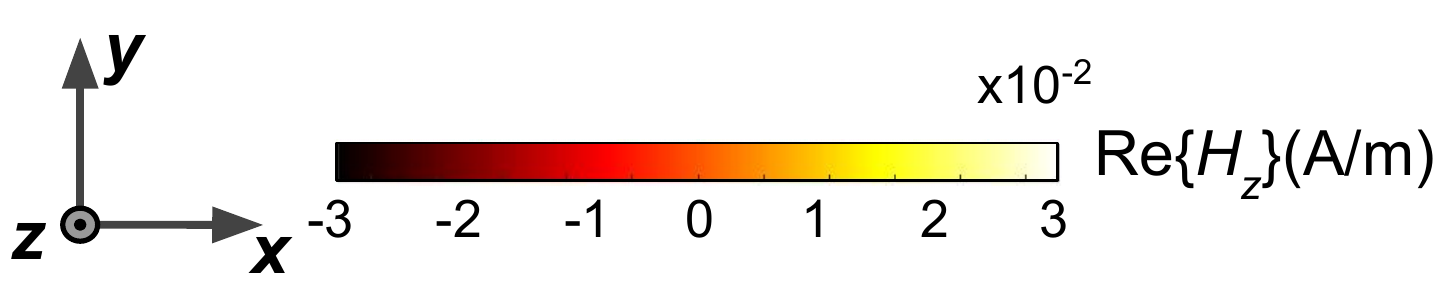}}               
              \end{gather*}  
     \begin{subfigure}[h]{0.15\textwidth}
      	\vspace{-0.2cm}   \includegraphics[width=\textwidth]{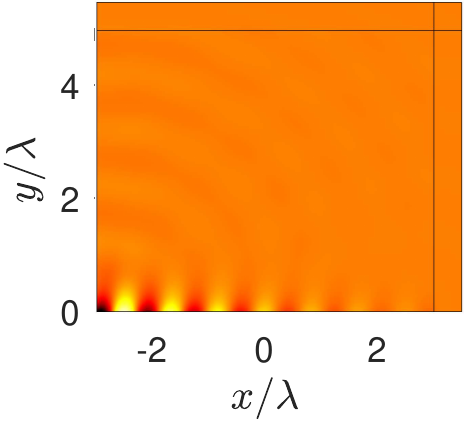}
        \caption{}
    \end{subfigure}
    		\begin{gather*}{
        \includegraphics[width=0.35\textwidth]{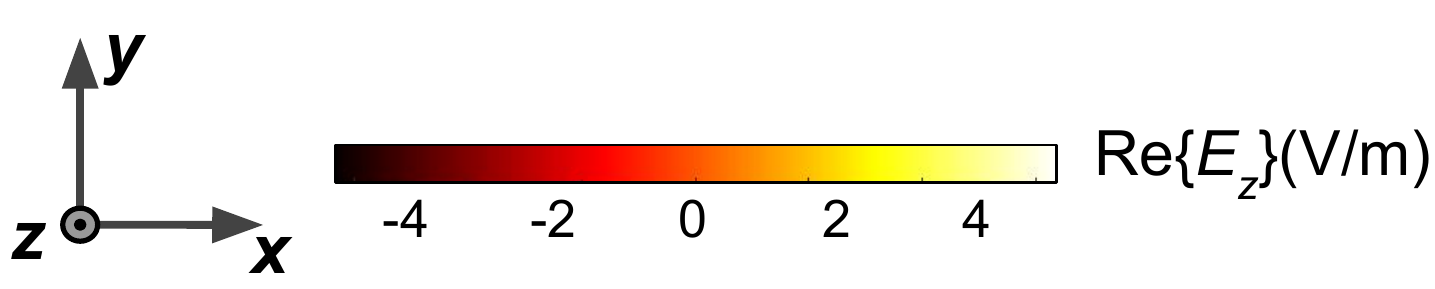}}               
              \end{gather*}  
	 \begin{subfigure}[h]{0.15\textwidth}
    	\vspace{-0.2cm}     \includegraphics[width=\textwidth]{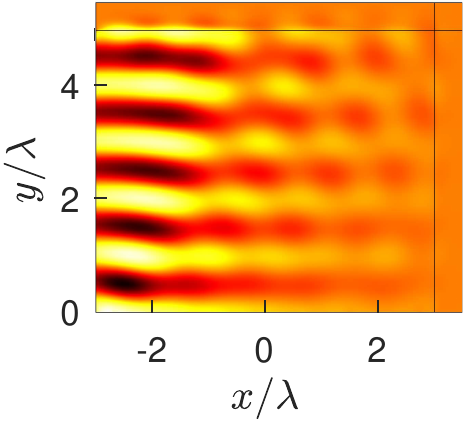}
        \caption{}
    \end{subfigure}\\  
\caption{\textcolor{black}{Numerically computed distributions of the tangential components of the total (a) magnetic field ${\rm{Re}}(H_z) $ and (b) electric field ${\rm Re}(E_z)$ }with the following wave parameters: ${\alpha _x} =  0.0844{k_0}$, $\alpha _y^{{\rm{sw}}} = 0.6748{k_0}$, $\beta _x^{{\rm{sw}}} = 1.2{k_0}$,$\beta _y^{{\rm{sw}}} = -0.15{k_0}$, $\beta _x^{{\rm{lw}}}=0$, $\beta _y^{{\rm{lw}}} =1.0036 {k_0}$, and $\alpha _y^{\rm lw}=0$.}
	\label{fig:4.2b}
\end{figure}
The surface wave propagation constants are chosen to be different from the previous case to ensure full leakage,  and they are equal to $\beta _x^{{\rm{sw}}} = 1.2{k_0}$,$\beta _y^{{\rm{sw}}} = -0.15{k_0}$. From  (\ref{eq:8}), (\ref{eq:9}) and (\ref{eq:15}) all other parameters of the waves can be found: ${\alpha _x} =  0.0844{k_0}$, $\alpha _y^{{\rm{sw}}} = 0.6748{k_0}$, $\beta _x^{{\rm{lw}}}=0$, $\beta _y^{{\rm{lw}}} =1.0036 {k_0}$, and $\alpha _y^{\rm lw}=0$. Conversion efficiency of the surface with the length of $6\lambda$ with these wave parameters is around $99\%$, while about $1\%$ of the energy is still not converted and carried by the surface wave.

\begin{figure}[H]
	\centering
	 \begin{subfigure}[h]{0.45\textwidth}
        \includegraphics[width=\textwidth]{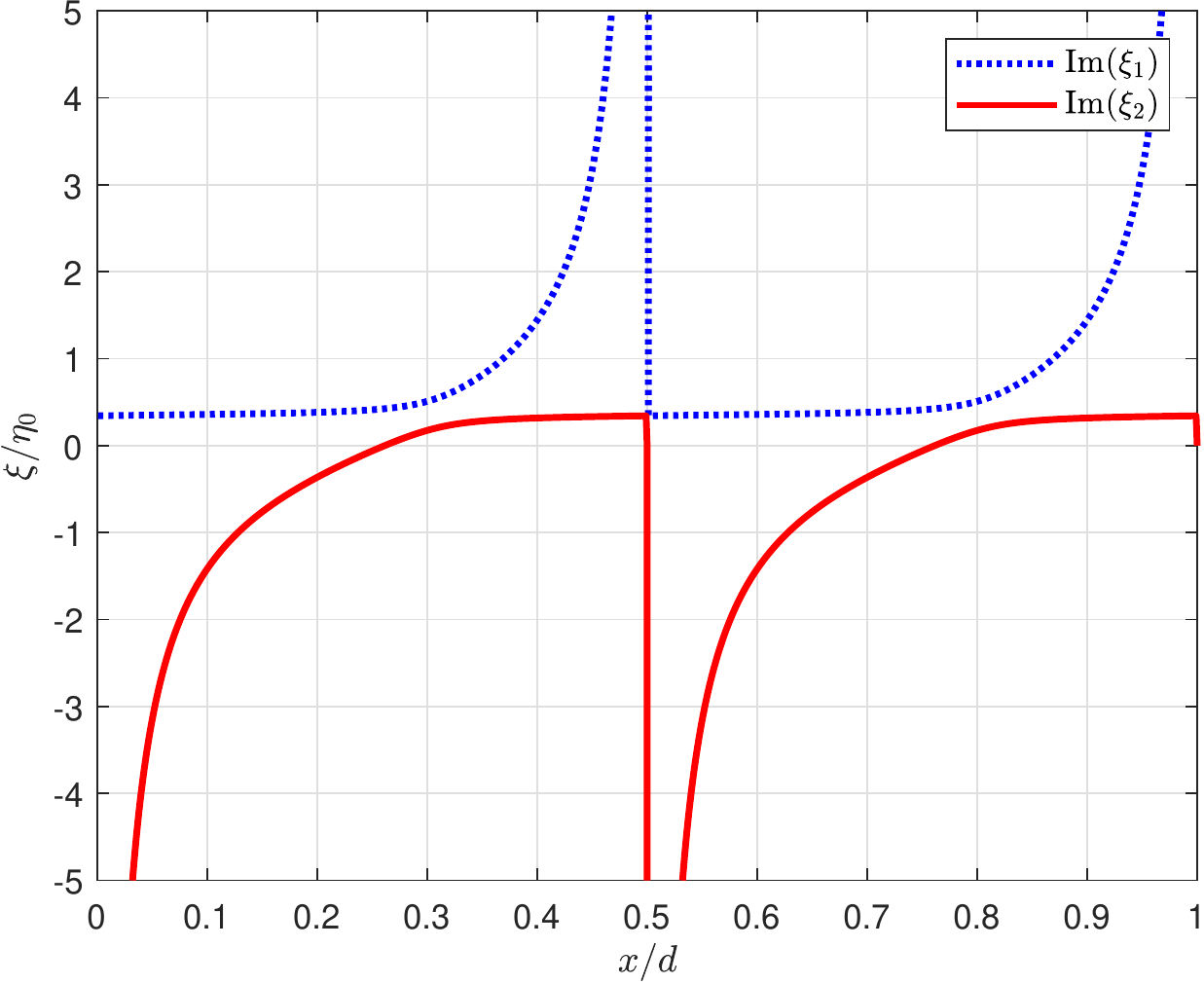}
        \caption{}
    \end{subfigure} 
  \begin{subfigure}[h]{0.45\textwidth}
    \includegraphics[width=\textwidth]{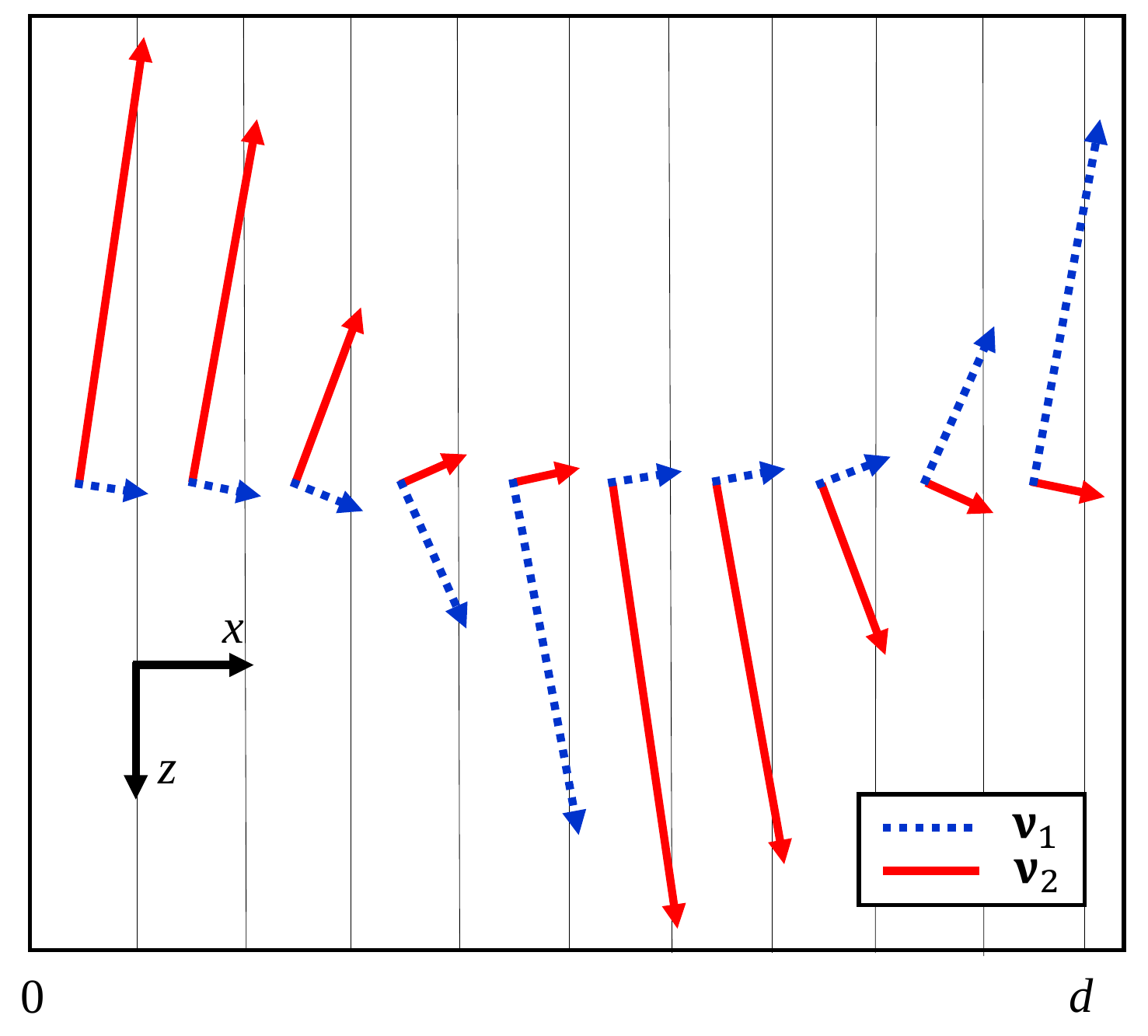}
        \caption{}
    \end{subfigure}
	\caption{(a) Eigenvalues $\xi_1$ and $\xi_2$ of the surface impedance matrix (\ref{eq:Zs}) and (b) eigenvectors ${\rm {\boldsymbol \nu}}_1$ and ${\rm {\boldsymbol \nu}}_2$, corresponding to $\xi_1$ and $\xi_2$, respectively .}
	\label{fig:eig}
\end{figure}

\subsection{\label{sec:4.4}{Suggestions for implementation by eigenvector analysis}}
One possibility for simplifying the realization of such metasurface is to transform the impedance matrix (\ref{eq:Zs}) in Cartesian coordinates into a diagonal one written in a new basis \cite{strang2016} . This transformation is equivalent to finding the eigenvalues (diagonal components) and the eigenvectors (new basis vectors) of the impedance matrix.
The eigenvalues can be easily calculated solving the equation $|\xi I-Z|=0$, where $\xi$ is an eigenvalue, $I$ is the identity matrix. Finally, eigenvalues are given by
\begin{equation}
\xi_{1,2}={1 \over 2} \left[  j(X_{xx}+X_{zz})\pm \sqrt{-4X_{xz}X_{zx}-(X_{xx}-X_{zz})^2}  \right].
\end{equation}
 %\noindent 
 Figure~\ref{fig:eig}(a) depicts two eigenvalues of the surface impedance along one period $d$.  The eigenvalues are purely imaginary and widely vary within a period spanning from capacitive to inductive values.

Eigenvectors ${\rm {\boldsymbol \nu}}_1$ and ${\rm {\boldsymbol \nu}}_2$ correspond to the found eigenvalues $\xi_1$ and $\xi_2$, and consist of two components (projections in the Cartesian coordinates) each, $\nu_{1,1}$, $\nu_{1,2}$ and $\nu_{2,1}$, $\nu_{2,2}$, respectively, that relate as
\begin{equation}
\nu_{1,2}=-{jX_{xx}-\xi_{1} \over jX_{xz}} \nu_{1,1}, \, \, \nu_{2,2}=-{jX_{xx}-\xi_{2} \over jX_{xz}} \nu_{2,1}.
\end{equation}
We plot Figure~\ref{fig:eig}(b) assuming $\nu_{1,1}=\nu_{2,1}$. The arrows show the direction for each of the eigenvectors as a function of the coordinate along the surface. Here, the surface is divided into ten equal cells per period and the eigenvalue from the middle of the cell is used for eigenvector calculation. 
%The eigenvectors show the rotation of the elements which represent specific eigenvalue relatively to $xz$-plane of the surface.

Eigenvalues and eigenvectors give the information of the resonant inclusions type and its reactive impedance value as well as its orientation on the surface plane. \textcolor{black}{The required surface reactance can be realized using arrays of properly oriented patches printed on a grounded dielectric substrate, as described in \cite{tcvetkova_prb2018}. } Design and practical realization of such converters is a scope of future research.\\

\section{Conclusions}
\label{sec:conclusion}

The problem of ideal conversion of a surface wave into a propagating plane wave  has been examined theoretically and numerically.
It is important to note, that the found exact solution of the boundary value problem for the reciprocal and lossless system is satisfied by one surface wave and one leaky wave only. 
Therefore, no higher-order Floquet modes of the asymptotic expansion which store the energy are excited near the radiating modulated reactive surface. 
It means that all the power carried by the surface wave is used for the creation of the leaky wave without storing reactive energy in higher order Floquet-modes. It is expected that the found exact solution represents a reference for minimal possible Q-factor for any leaky wave antenna, thus establishing a bound of maximum bandwidth. \textcolor{black}{In this scenario, the reactive energy stored in  the meta-atoms used for practical realization of the metasurface would be a limiting factor for the bandwidth.}
This investigation of maximum bandwidth will be the subject of an incoming paper.

\section*{Acknowledgment}
This work was supported in part by the Academy of Finland (project 287894), Nokia Foundation (project 201810155 and 201910452), HPY Research Foundation and European Space Agency.

\clearpage
\appendix
\section{General surface impedance formulation}
  \label{ap:1}
The general surface impedance formulation for oblique incidence. In the present paper the particular case of the normal propagating wave radiation is considered.
\begin{strip}
\begin{align}
X_{xx}&=\displaystyle {{\eta_0 \over k_0}(\alpha_y^{\rm sw}-\beta_y^{\rm sw}\cot{\beta_x^{\rm sw}x})}+ \nonumber \\
\displaystyle &\hspace{1cm} {{\eta_0 \over k_0} {\beta_y^{\rm sw} \over \sin{\beta_x^{\rm sw}x}} { \alpha_y^{\rm lw}\cos{\beta_x^{\rm lw}x}+\beta_y^{\rm lw}\sin{\beta_x^{\rm lw}x} \over \alpha_y^{\rm lw}(\cos{\beta_x^{\rm lw}x}\cos{\beta_x^{\rm sw}x} +\sin{\beta_x^{\rm lw}x}\sin{\beta_x^{\rm sw}x}           )+\beta_y^{\rm lw}(\sin{\beta_x^{\rm lw}x}\cos{\beta_x^{\rm sw}x} -\cos{\beta_x^{\rm lw}x}\sin{\beta_x^{\rm sw}x}   )}} \\
X_{xz}&=X_{zx}=\displaystyle \sqrt{-{\beta_y^{\rm sw} \over \beta_y^{\rm lw}}} \displaystyle \left[{\beta_y^{\rm lw} \eta_0 \over \beta_y^{\rm lw}(\cos{\beta_x^{\rm lw} x}\sin{\beta_x^{\rm sw} x}-\sin{\beta_x^{\rm lw} x}\cos{\beta_x^{\rm sw} x})- \alpha_y^{\rm lw}(\sin{\beta_x^{\rm lw} x}\sin{\beta_x^{\rm sw} x}+\cos{\beta_x^{\rm lw} x}\cos{\beta_x^{\rm sw} x}) } \right]\\
X_{zz}&=\displaystyle {\eta_0 k_0 (\cos{\beta_x^{\rm lw}x}\cos{\beta_x^{\rm sw}x}+\sin{\beta_x^{\rm lw}x}\sin{\beta_x^{\rm sw}x})  \over  \beta_y^{\rm lw}(\sin{\beta_x^{\rm sw}x} \, \cos{\beta_x^{\rm lw}x}-\sin{\beta_x^{\rm lw}x}\cos{\beta_x^{\rm sw}x})-\alpha_y^{\rm lw} (\sin{\beta_x^{\rm sw}x} \, \sin{\beta_x^{\rm lw}}+\cos{\beta_x^{\rm sw}x}\cos{\beta_x^{\rm lw}x})    }
\end{align}
\end{strip} 

%\bibliography{IEEEabrv,references}
%\bibliographystyle{IEEEtr}
%\bibliography{references}
	
\bibliography{IEEEabrv,references}
\bibliographystyle{IEEEtran}

\begin{IEEEbiography}[{\includegraphics[width=1in,height=1.25in,clip,keepaspectratio]{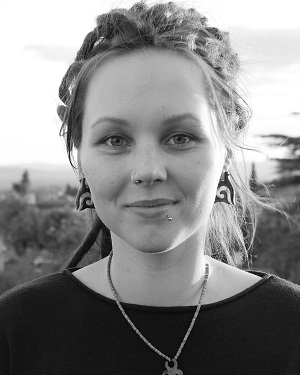}}]{Svetlana Tcvetkova} (M'18) 
was born in Vyshny Volochyok, Russia, in 1991. She received the B.Sc. and M.Sc. degrees in technical physics from the Peter the Great St. Petersburg Polytechnic University, Russia, in 2013 and 2015, respectively, as well as the M.Sc. (Tech.) degree in computational engineering from Lappeenranta University of Technology, Finland, in 2015. 

She is currently working towards the D.Sc. degree with the Department of Electronics and Nanoengineering, Aalto University, in the Group of Prof. S. Tretyakov.
In 2017, she was a Visiting Student Researcher in
the Group of Prof. S. Maci, Siena University, Siena, Italy.
Her main research interests include antennas, microwave measurements, and reflection/transmission properties of metasurfaces.

\end{IEEEbiography}

\begin{IEEEbiography}[{\includegraphics[width=1in,height=1.25in,clip,keepaspectratio]{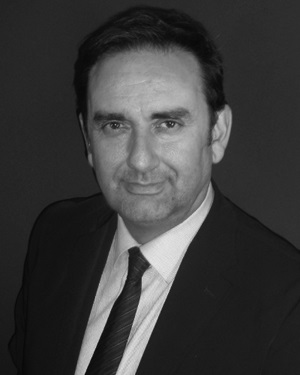}}]{Stefano Maci} (S'98--F'04) received the Laurea degree (cum laude) in electronic engineering from the
University of Florence, Italy.
Since 1998, he has been with the University of
Siena, Italy, where he is presently a Full Professor,
Director of the Ph.D. School of Engineering and
Head of the Laboratory of Electromagnetic Applications (LEA). His research interests include EM
theory, antennas, high-frequency methods, computational electromagnetics, and metamaterials. He was
a coauthor of an Incremental Theory of Diffraction
for the description of a wide class of electromagnetic scattering phenomena at
high frequency, and of a diffraction theory for the analysis of large truncated
periodic structures. He was responsible and the international coordinator of
several research projects funded by the European Union (EU), by the European
Space Agency (ESA-ESTEC), by the European Defence Agency, and by
various European industries. He was the founder and is currently the Director
of the European School of Antennas (ESoA), a post-graduate school that
comprises 30 courses on antennas, propagation, and EM modelling though by
150 teachers coming from 30 European research centers. He is the principal
author or coauthor of 110 papers published in international journals, (among
which 70 are in IEEE journals), 10 book chapters, and about 350 papers in
proceedings of international conferences.
Prof. Maci is member of the Finmeccanica. He was an Associate Editor of
the IEEE TRANSACTIONS ON ELECTROMAGNETIC COMPATIBILITY, twice Guest
Editor and an Associate Editor of IEEE TRANSACTION ON ANTENNAS AND
PROPAGATION (IEEE-TAP). He is presently a member of the IEEE AP-Society
AdCom, a member of the Board of Directors of the European Association on
Antennas and Propagation (EuRAAP), a member of the Executive Team of the
IET Antennas and Propagation Network, a member of the Technical Advisory
Board of the URSI Commission B, and a member of the Italian Society of
Electromagnetism. He was the recipient of several national and international
prizes and best paper awards.
\end{IEEEbiography}

\begin{IEEEbiography}[{\includegraphics[width=1in,height=1.25in,clip,keepaspectratio]{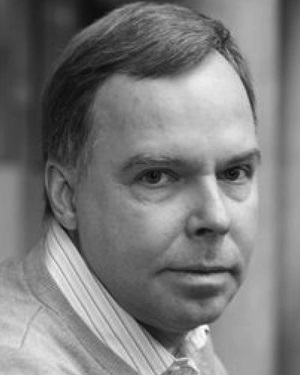}}]{Sergei A. Tretyakov} (M'92–-SM'98–-F'08) received the Dipl.Ing.-Physicist, Ph.D., and D.Sc. degrees
from the Saint Petersburg State Technical University,
Saint Petersburg, Russia, in 1980, 1987, and 1995,
respectively, all in radiophysics.
From 1980 to 2000, he was with the Radiophysics
Department, Saint Petersburg State Technical University. He is currently a Professor of radio science
with the Department of Electronics and Nanoengineering, Aalto University, Espoo, Finland. He has
authored or co-authored five research monographs
and over 270 journal papers. His current research interests include electromagnetic field theory, complex media electromagnetics, metamaterials, and
microwave engineering.
Dr. Tretyakov served as the President for the Virtual Institute for Artificial
Electromagnetic Materials and Metamaterials (Metamorphose VI), the General
Chair for the International Congress Series on Advanced Electromagnetic
Materials in Microwaves and Optics (Metamaterials) from 2007 to 2013, and
the Chairman for the Saint Petersburg IEEE Electron Devices/Microwave Theory and the Techniques/Antennas and Propagation Chapter from 1995 to 1998.
\end{IEEEbiography}

\end{document}